\documentclass[twocolumn]{book}
\usepackage[dvips]{graphicx,color}
\usepackage{makeidx,universe}
\usepackage{epsfig}
\usepackage{wrapfig,rotating}
\usepackage{amssymb}
\usepackage{amsmath}

\makeauthorindex

\newcommand{\PO}{\rm l \! P }
\newcommand{\gv}{\gamma^\star}
\newcommand{\RO}{\rm l \! R }

\newcommand{\gvp}{\gamma^\star p}
\newcommand{\xpom}{x_{\PO} }

\newcommand{\be}{\begin{equation}}
\newcommand{\ee}{\end{equation}}

\BookTitle{Proceedings of the XXIX PHYSICS IN COLLISION}

\CopyRight{\copyright 2009 by Universal Academy Press, Inc.}

\begin{document}
\pagenumbering{arabic}

\chapter{%
{\LARGE \sf
Diffraction and its QCD interpretation } \\
{\normalsize \bf 
Laurent SCHOEFFEL$^1$ } \\
{\small \it \vspace{-.5\baselineskip}
(1) CEA Saclay/Irfu-SPP, 91191 Gif-sur-Yvette, France
}
}

  \baselineskip=10pt 
  \parindent=10pt    

\section*{Abstract} 

The most important results on hadronic diffractive phenomena 
obtained at HERA and Tevtaron are reviewed and 
new issues in nucleon tomography are discussed.
Some challenges for understanding
diffraction at the LHC, including the discovering of the Higgs boson, 
are outlined.


\section{Experimental diffraction at HERA}

Between 1992 and 2007, the 
HERA accelerator provided $ep$ collisions at center
of mass energies beyond $300 \ {\rm GeV}$
at the interaction points of the H1 and ZEUS experiments. 
Perhaps the most interesting results to emerge relate to
the newly accessed field of 
perturbative strong interaction physics at low Bjorken-$x$ ($x_{Bj}$),
where parton densities become extremely large.
Questions arise as to how and where non-linear dynamics tame
the parton density growth
and challenging features 
are observed. Central to this low $x_{Bj}$ physics landscape 
is a high rate of diffractive
processes, in which a colorless exchange takes place and
the proton remains intact. 
Indeed,
one of the most important experimental result from the DESY $ep$ collider HERA
is the observation of a significant fraction of events in Deep Inelastic Scattering (DIS)
with a large rapidity gap (LRG) between the scattered proton, which remains intact,
and the rest of the final system. This fraction corresponds to about 10\% of the DIS   data
at $Q^2=10$ GeV$^2$.

\begin{figure}[htbp]
\begin{center}
\psfig{figure=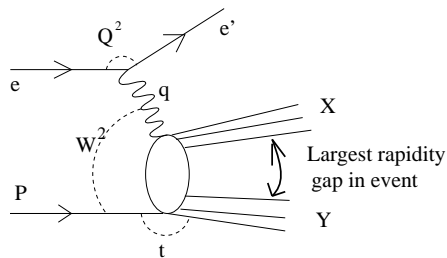,width=0.33\textwidth,angle=0}
\end{center}
\caption{Picture of the process $ep \rightarrow eXY$. The
hadronic final state is composed of two distinct systems $X$ and
$Y$, which are separated by the largest interval 
in rapidity between final state hadrons.}
\label{difproc}
\end{figure}

In DIS, such events are not expected in such abundance, since large gaps are exponentially
suppressed due to color string formation between the proton remnant and the scattered partons.
Events
are of the type $ep \rightarrow eXp$, where the final state proton
carries more than $95$ \% of the proton beam energy. 
A photon of virtuality $Q^2$, coupled to the electron (or positron),
undergoes a strong interaction with the proton (or one of its 
low-mass excited states $Y$) to form a hadronic final state
system $X$ of mass $M_X$ separated by a LRG
from the leading proton (see Fig. \ref{difproc}). 
These events are called diffractive.
In such a reaction, $ep \rightarrow eXp$,
no net quantum number is exchanged and 
  the longitudinal momentum fraction $1-x_{\PO}$  
  is lost by the proton. Thus, the mongitudinal momentum $\xpom P$ is transfered 
to the system $X$. In addition to the standard DIS kinematic variables and $\xpom$, 
a diffractive event is also often 
characterized by the variable $\beta={x_{Bj}}/{x_{\PO}}$, which takes a simple 
interpretation in the parton model discussed in the following.
Comparisons
with hard diffraction in proton-(anti)proton scattering have
also improved our knowledge of absorptive 
and underlying event effects in which the 
diffractive signature may be obscured by multiple interactions in the
same event. In addition to their fundamental
interest in their own right, these issues are highly relevant
to the modeling of chromodynamics at the LHC.  

Experimentally, for
a diffractive DIS event, $ep\rightarrow eXp$, the dissociating
particle is the virtual photon emitted by the electron. The final
state consists of the scattered electron and hadrons which populate
the photon fragmentation region. The proton is scattered in the
direction of the initial beam proton with little change in 
momentum and angle. In particular, we detect no hadronic activity in the
direction of the proton flight, as the proton remains intact 
in the diffractive process. On the contrary, for a standard DIS event,
the proton is destroyed in the reaction and the flow of hadronic clusters
is clearly visible in the proton fragmentation region (forward part of the detector).

The experimental selection of diffractive events in DIS proceeds in two steps.  
Events are first selected based on the presence of the scattered
electron in the detector. Then, for
the diffractive selection itself, three different methods have been used at
HERA: 
\begin{enumerate} 
\item A reconstructed proton track is required in the leading (or forward) proton 
spectrometer (LPS for ZEUS or FPS for H1) with a fraction of the initial proton momentum
$x_L>0.97$. Indeed,  the cleanest selection
of diffractive events with photon dissociation is based on the
presence of a leading proton in the final state. By leading proton we
mean a proton which carries a large fraction of the initial beam
proton momentum. This is the cleanest way to select diffractive events, but
the disadvantage is a reduced kinematic coverage.
\item The hadronic system $X$ measured in the central detector is 
required to be separated by a large rapidity gap from the rest of the
hadronic final state. This is a very efficient way to select diffractive events
in a large kinematic domain, close to the standard DIS one. The prejudice is 
a large background as discussed in the following.
\item The diffractive contribution is identified as
the excess of events at small $M_X$ above the exponential fall-off of
the non-diffractive contribution with decreasing $\ln M^2_X$. The exponential fall-off, expected in
QCD, permits the subtraction of the non-diffractive
contribution and therefore the extraction of the diffractive
contribution without assuming the precise $M_X$ dependence of the
latter. This is also a very efficient way to select diffractive events
in a large kinematic domain.
\end{enumerate}

\begin{figure}[tbp]
\begin{center}
\psfig{figure=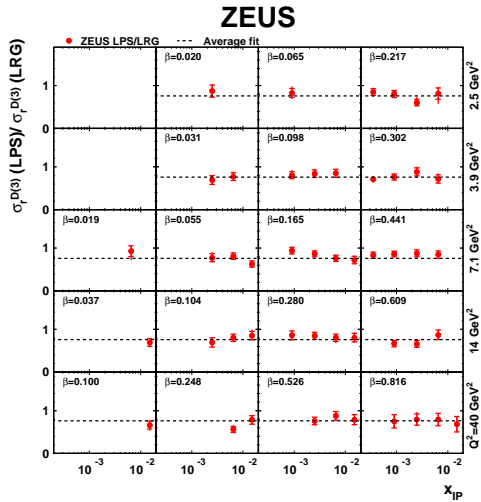,width=0.4\textwidth,angle=0}
\end{center}
\caption{Ratio of the diffractive cross sections, as obtained with the LPS and
the LRG experimental techniques. The lines indicate the average value of the ratio,
which is about 0.86. It implies that the LRG sample contains about
24\% of proton dissociation  events, corresponding to processes like $ep \rightarrow eXY$,
where $M_Y<2.3$ GeV. This fraction is approximately the same for H1 data (of course in the same $M_Y$ range).}
\label{lpsoverlrg}
\end{figure}
\begin{figure}[tbp]
\begin{center}
\psfig{figure=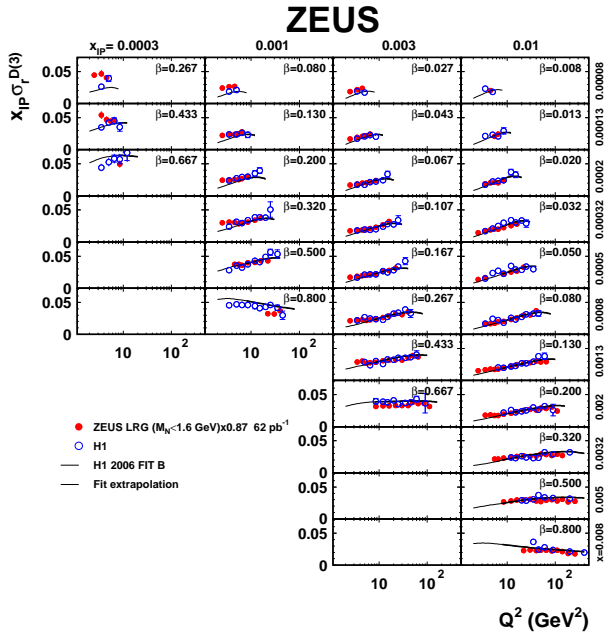,width=0.5\textwidth,angle=0}
\end{center}
\caption{The diffractive cross sections obtained with the LRG method by the H1 and ZEUS experiments.
The ZEUS values have been rescaled (down) by a global factor of 13 \%. This value is compatible with the normalisation uncertainty
of this sample. }
\label{datah1zeus}
\end{figure}
\begin{figure}[tbp]
\begin{center}
\psfig{figure=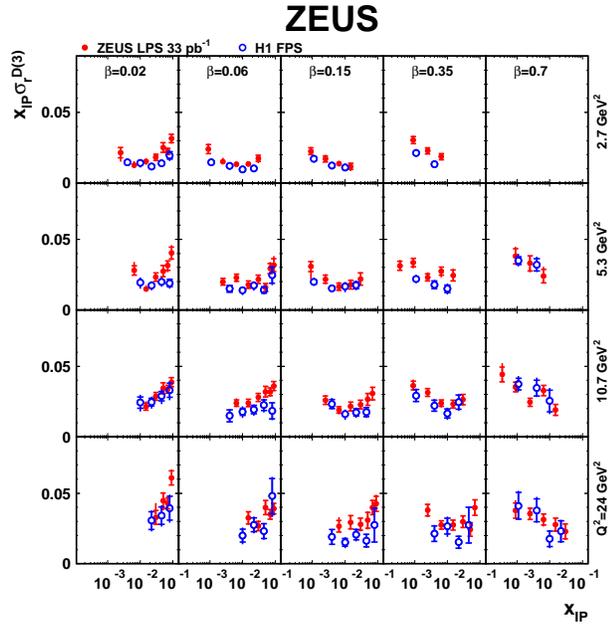,width=0.5\textwidth,angle=0}
\end{center}
\caption{The diffractive cross section obtained with the FPS (or LPS) method by the H1 and ZEUS experiments,
where the proton is tagged. The ZEUS measurements are above H1 by a global factor of about 10\%.}
\label{datah1zeuslps}
\end{figure}

\noindent
Extensive measurements of diffractive DIS cross sections have been made by both
the ZEUS and H1 collaborations at HERA,
using different experimental techniques \cite{f2d97,marta}. 
Of course, the comparison of these techniques provides a rich
source of information to get a better understanding of the experimental gains
and prejudices of those techniques.
These last published set of data  \cite{marta} contain
five to seven times more statistics than in preceding publications of diffractive
cross sections, and thus opens the way to new developments in data/models comparisons.
A first relative control of the  data samples is shown in Fig. \ref{lpsoverlrg}, where the
ratio of the diffractive cross sections is displayed,
as obtained with the LPS and
the LRG experimental techniques. The mean value of the ratio of $0.86$ indicates that
the LRG sample contains about 24\% of proton-dissociation background, which is not
present in the LPS sample. This background corresponds to events like
$ep \rightarrow e X Y$, where $Y$ is a low-mass excited state of the proton (with
$M_Y < 2.3$ GeV). 
It is obviously not present in the LPS analysis which  can select specifically a proton
in the final state.   This is the main background in the LRG analysis. Due to a lack
of knowledge of this background, it causes a large normalization uncertainty of 10  to 15 \% for the
cross sections extracted from the LRG analysis.
We can then compare the results obtained by the H1 and ZEUS experiments for diffractive
cross sections (in Fig. \ref{datah1zeus}), using the LRG method.
A good compatibility of both data sets is observed, after rescaling the ZEUS points by 
a global factor of 13\%. This factor is compatible with the normalization uncertainty described above.
We can also compare the results obtained by the H1 and ZEUS experiments (in Fig. \ref{datah1zeuslps}),
 using the tagged proton  method (LPS for ZEUS and FPS for H1).
In this case, there is no proton dissociation background and the diffractive sample
is expected to be clean. It gives a good reference to compare both experiments. A global
normalization difference of about 10\% can be observed in Fig. \ref{datah1zeuslps},
which can be studied with more data. It remains compatible with the normalization
uncertainty for this tagged proton  sample.
It is interesting to note that the ZEUS measurements are globally above the H1 data by 
 about 10\% for both techniques, tagged proton or LRG.
The important message at this level is not only the observation of differences
as illustrated in Fig.  \ref{datah1zeus} and   \ref{datah1zeuslps},
but the opportunity opened with the large statistics provided by the ZEUS measurements. 
Understanding discrepancies
between data sets is part of the experimental challenge. It certainly needs
 analysis of new data sets from the H1 experiment. However, already at the present
level, much can be done with existing data for the understanding of  diffraction at HERA.

\section{Soft physics at the proton vertex}
\label{soft}

\begin{figure}[htbp]
            \includegraphics[width=0.45\textwidth]{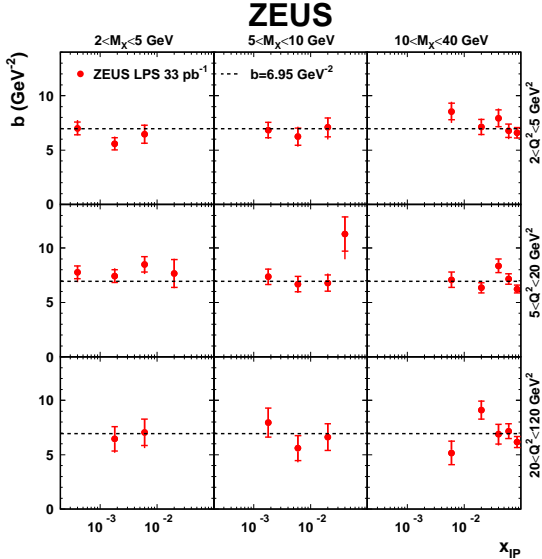}            
            \includegraphics[width=0.45\textwidth]{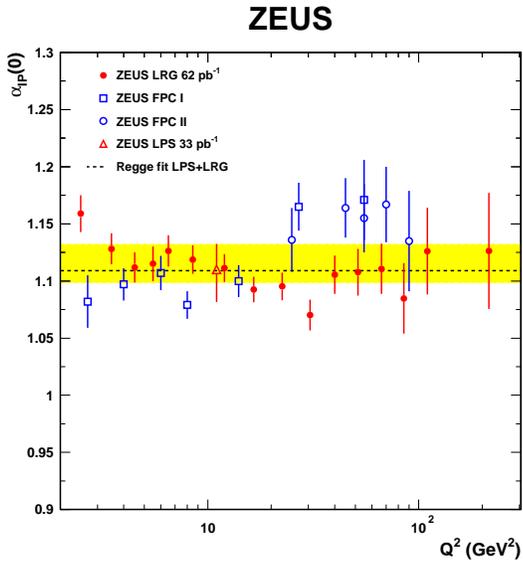}
\caption{(a) -top- Measurements of the exponential $t$ slope from ZEUS
LPS data, shown as a function of $Q^2$, $x_{I\!\!P}$ and $M_X$.
(b) -bottom- ZEUS extractions of the effective pomeron intercept describing
the $x_{I\!\!P}$ dependence of diffractive DIS data at different $Q^2$ 
values.}
\label{regge}
\end{figure}

To good approximation, LRG and LPS 
data show that 
diffractive DIS data satisfy a proton vertex 
factorization,
whereby 
the dependences on variables which describe the scattered 
proton ($x_{I\!\!P}$, $t$) factorize from those
describing the hard partonic interaction ($Q^2$, $\beta$).
For example, 
the slope parameter $b$, extracted 
by fitting the $t$ distribution to the form
${\rm d} \sigma / {\rm d} t \propto e^{b t}$, is shown as a function
of diffractive DIS (DDIS) kinematic variables in Fig.~\ref{regge}a. 
There are no significant variations
from the average value of $b \simeq 7 \ {\rm GeV^{-2}}$ anywhere in the
studied range.   
The measured value of $b$ is significantly larger than
that from `hard' 
exclusive vector meson production ($ep \rightarrow eVp$). 
It is 
characteristic of an interaction region 
of spatial extent considerably larger than the proton radius, 
indicating that the dominant feature of DDIS is the probing with
the virtual photon of
non-perturbative exchanges similar to the Pomeron 
of soft hadronic physics. 

Fig.~\ref{regge}b shows the $Q^2$ dependence of the effective
Pomeron intercept $\alpha_{I\!\!P} (0)$, which is extracted from 
the $x_{I\!\!P}$ dependence of the data. 
No significant
dependence on $Q^2$ is observed, again compatible with
proton vertex factorization. 

The intercept 
of the effective Pomeron trajectory
is consistent within errors with the soft Pomeron results 
from fits to total cross sections and
soft diffractive data.
Although larger effective intercepts 
have been measured in hard vector meson production, 
no deviations with either $Q^2$ or $\beta$ have yet been observed
in inclusive diffractive DIS.

\section{Diffractive PDFs  at HERA}
\begin{figure}[htbp]
\begin{center}
\includegraphics[width=8cm,height=8.5cm]{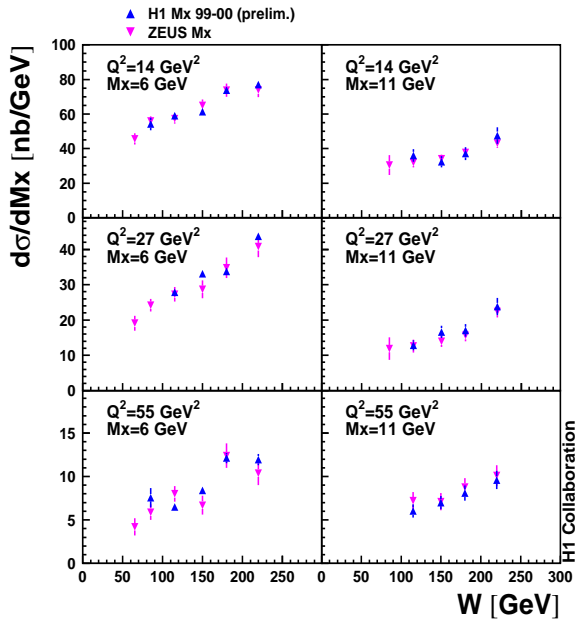}
\caption{Cross sections of the diffractive process $\gamma^* p \rightarrow p' X$, 
differential in the mass of the diffractively produced hadronic system $X$ ($M_X$),
are presented as a function of the center-of-mass energy of the $\gamma^*p$ system $W$.
Measurements at different values of the virtuality
$Q^2$ of the exchanged photon are displayed. We observe a behavior of the form $\sim W^{ 0.6}$  for the diffractive cross section, 
compatible with the dependence expected
for a hard process. 
}
\label{figdata}
\end{center}
\vspace{-0.5cm}
\end{figure}

\begin{figure}[htbp]
\begin{center}
\psfig{figure=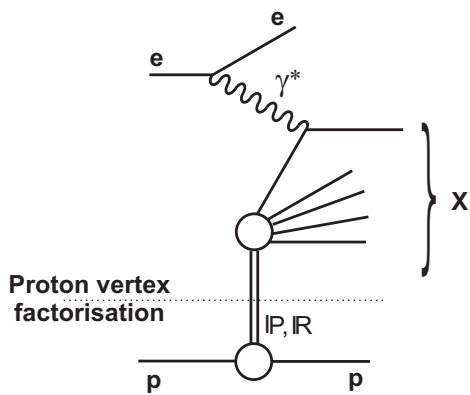,width=0.35\textwidth,angle=0}
\end{center}
\caption{Schematic diagram of a diffractive process.
Events with a diffractive topology can be studied  in terms of
the Pomeron trajectory exchanged between the proton and the virtual photon.
}
\label{pomeron}
\end{figure}

In order to compare diffractive data with perturbative QCD models, or parton-driven models,
the first step is to show that 
the diffractive cross section
shows a hard dependence in the center-of-mass energy $W$ of the $\gamma^*p$ system.
In Fig. \ref{figdata}, we observe a behavior of the form $\sim W^{ 0.6}$ , 
compatible with the dependence expected
for a hard process. This  observation is obviously 
the key to allow further studies of the diffractive process in the context of
perturbative QCD \cite{collins}.
Events with the diffractive topology can be studied  in terms of
Pomeron trajectory exchanged between the proton and the virtual photon.
In this view, these events result from a color-singlet exchange
between the diffractively dissociated virtual photon and the proton (see Fig. \ref{pomeron}). 

\begin{figure}[htbp]
\begin{center}
\includegraphics[width=0.3\textwidth]{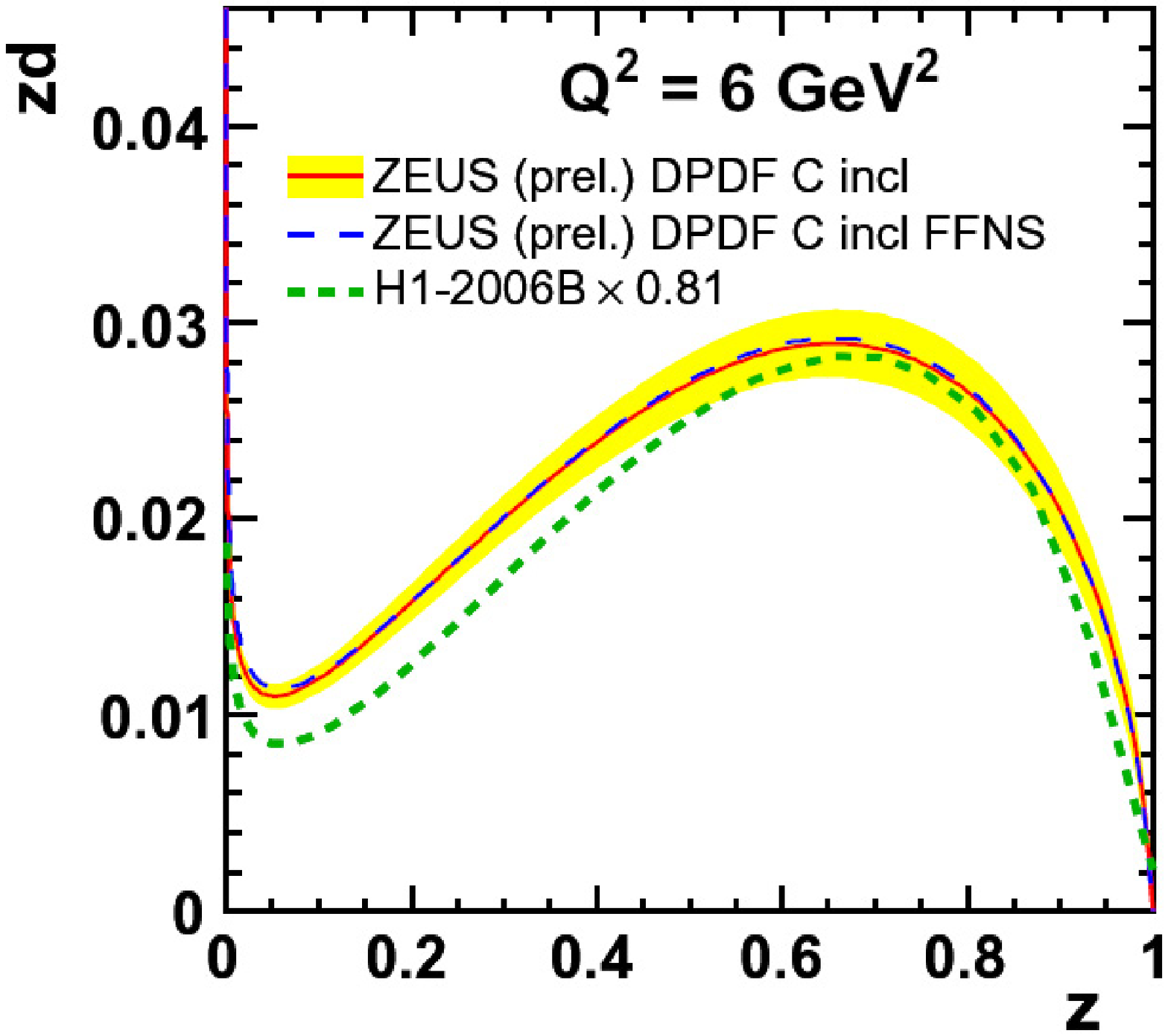}
\includegraphics[width=0.3\textwidth]{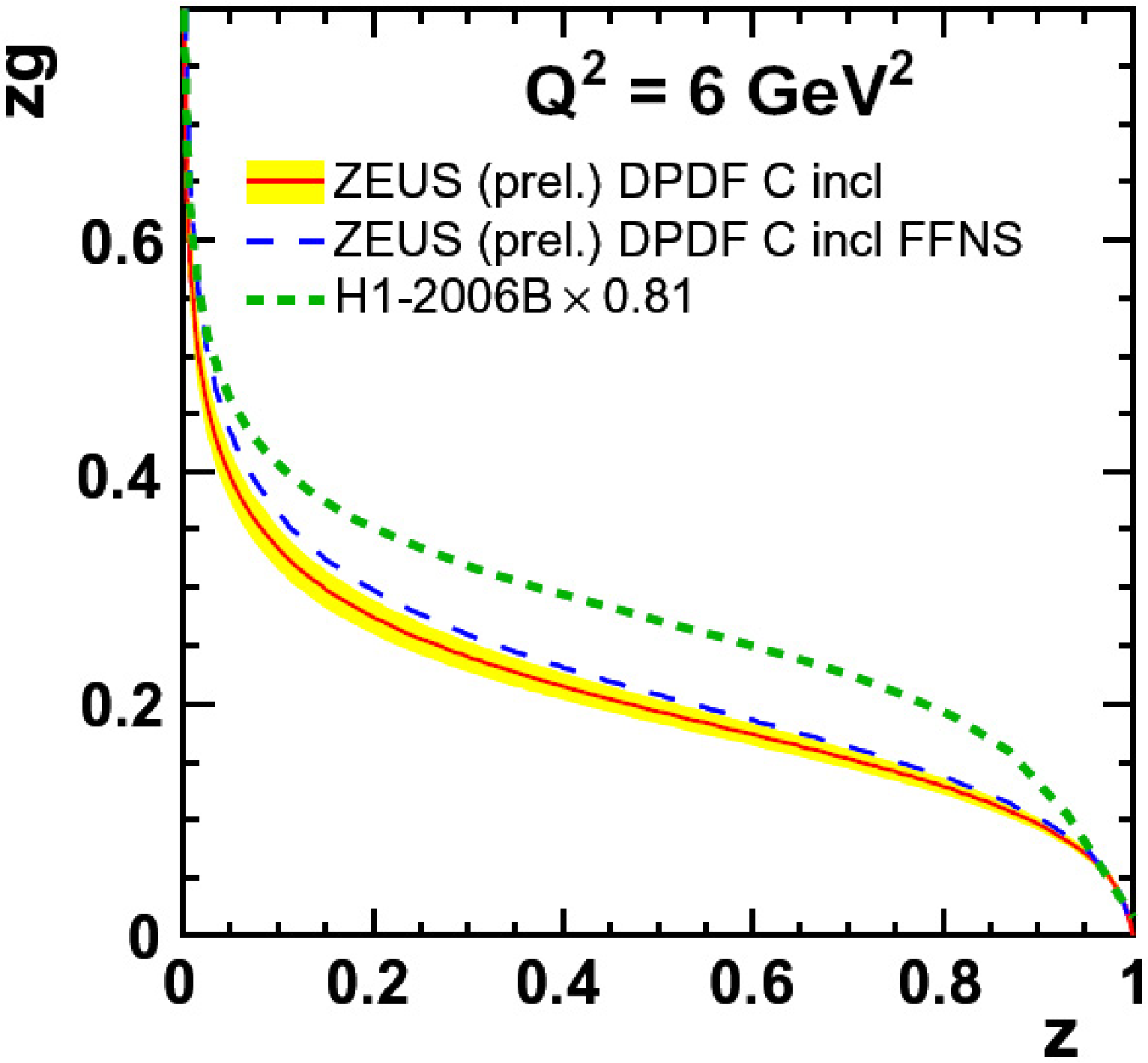}
\caption{ZEUS down quark (one sixth of the total quark $+$ 
antiquark) and gluon densities
as a function of generalised momentum fraction $z$ at 
$Q^2 = 6 \ {\rm GeV^2}$ \cite{zeus:diffqcd}. 
Two heavy flavour schemes are shown, 
as well as H1 results corrected
for proton dissociation with a factor of 0.81.}
\label{dpdfsh1}
\end{center}
\end{figure}

A diffractive
structure function $F_2^{D(3)}$ 
can then be defined as a sum of two factorized 
contributions, corresponding to a Pomeron and secondary Reggeon trajectories: 
$$
F_2^{D(3)}(Q^2,\beta,x_{\PO})=
f_{\PO / p} (x_{\PO}) F_2^{D(\PO)} (Q^2,\beta) 
$$
$$
+ f_{\RO / p} (x_{\PO}) F_2^{D(\RO)} (Q^2,\beta),
$$
where $f_{\PO / p} (x_{\PO})$ is the Pomeron flux. It depends only on $\xpom$,
once integrated over $t$, and
$F_2^{D(\PO)}$ can be interpreted as the Pomeron structure function,
depending on $\beta$ and $Q^2$.
The other function,
$F_2^{D(\RO)}$, is an effective Reggeon structure function
taking into account various secondary Regge contributions which can not be 
separated.
The Pomeron and Reggeon fluxes are assumed to follow a Regge behavior with  
linear
trajectories $\alpha_{\PO,\RO}(t)=\alpha_{\PO,\RO}(0)+\alpha^{'}_{\PO,\RO} t$, 
such that
\begin{equation}
f_{{\PO} / p,{\RO} / p} (x_{\PO})= \int^{t_{min}}_{t_{cut}} 
\frac{e^{B_{{\PO},{\RO}}t}}
{x_{\PO}^{2 \alpha_{{\PO},{\RO}}(t) -1}} {\rm d} t ,
\label{flux}
\end{equation}
where $|t_{min}|$ is the minimum kinematically allowed value of $|t|$ and
$t_{cut}=-1$ GeV$^2$ is the limit of the measurement. 
We take
$\alpha^{'}_{\PO}=0.06$ GeV$^{-2}$, 
$\alpha^{'}_{\RO}=0.30$ GeV$^{-2}$,
$B_{\PO}=5.5$ GeV$^{-2}$ and $B_{\RO}=1.6$ GeV$^{-2}$. 
The Pomeron
intercept $\alpha_{\PO}(0)$ is left as a free parameter in the QCD fit
and $\alpha_{\RO}(0)$ is fixed to $0.50$.

\begin{figure}[!]
\begin{center}
\psfig{figure=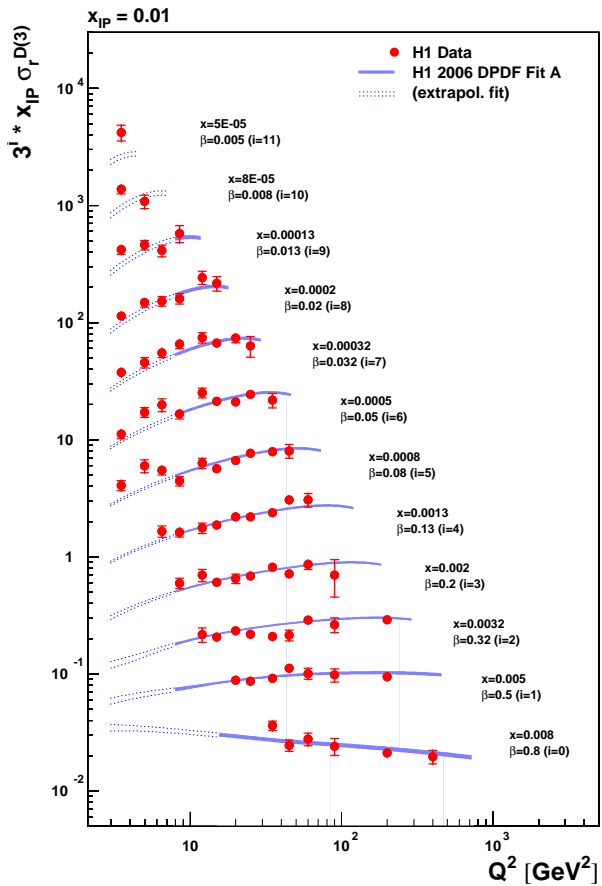,width=0.45\textwidth,angle=0}
\end{center}
\caption{Scaling violations for H1 diffractive cross sections for one value of $\xpom$ ($\xpom=0.01$)
and a large range of  $\beta$ values, from low  ($<0.01$) to large values ($> 0.5$).}
\label{scalingviol}
\end{figure}

The next step is then to model the Pomeron structure function $F_2^{D(\PO)}$
\cite{f2d97,lolo1,zeus:diffqcd}.
Among the most popular models, the one based on a point-like structure of
the Pomeron has been studied extensively 
using a non-perturbative input supplemented by a perturbative QCD evolution equations
\cite{lolo1,zeus:diffqcd}.
In this formulation, it is assumed that the exchanged object, the Pomeron, 
is a color-singlet quasi-particle whose structure is probed in the
DIS process. 
As for standard DIS,   diffractive parton distributions  
related to the Pomeron can be derived from QCD fits to diffractive cross sections.
The procedure is standard: we assign parton distribution functions to the Pomeron
parametrized in terms of non-perturbative input
distributions at some low scale $Q_0^2$. The quark flavor singlet distribution
($z{ {S}}(z,Q^2)=u+\bar{u}+d+\bar{d}+s+\bar{s}$)
and the gluon distribution ($z{\it {G}}(z,Q^2)$) are parametrized 
at this initial scale
$Q^2_0$, where $z=x_{i/I\!\!P}$ is the fractional momentum of the Pomeron carried by
the struck parton. Functions $z{\it{S}}$ and $z{\it{G}}$  
are evolved to higher $Q^2$ using the
next-to-leading order DGLAP evolution equations.
For the structure of the sub-leading Reggeon trajectory,
the pion structure function 
\cite{owens} is assumed
with a free global normalization to be determined by the data. 
Diffractive PDFs (DPDFs) extracted from 
H1 and ZEUS data are shown in Fig. \ref{dpdfsh1} 
\cite{f2d97,lolo1,zeus:diffqcd}. 
We observe that some differences in the data are reflected in the DPDFs, but some
basic features are common for all data sets and the resulting DPDFs. Firstly, the gluon density
is larger than the sea quark density, which means that the major fraction of the momentum
(about 70\%) is carried by the gluon for a typical value of $Q^2=10$ GeV$^2$. Secondly, we observe 
that the gluon density is quite large at large $\beta$, with a large uncertainty, which means that
we expect positive scaling violations still at large values of $\beta$. This is shown in Fig. \ref{scalingviol}.
We note that even at large values of $\beta \sim 0.5$, the scaling violations are still positive,
as discussed above. The strength of the DPDFs approach is to give a natural interpretation of this
basic observation and to describe properly the $Q^2$ evolution of the cross sections.
Other approaches are also well designed to describe all features of the data \cite{plb}, but this
is another story.
The near future of the study of DPFDs is to combine all existing data and check their
compatibility with respect to the QCD fit technique. If this is verified, a new global analysis
can be followed to get the most complete understanding of DPDFs \cite{lolo1}.

\section{DPDFs and implications for the LHC}

\begin{figure}[!]
\begin{center}
\epsfig{file=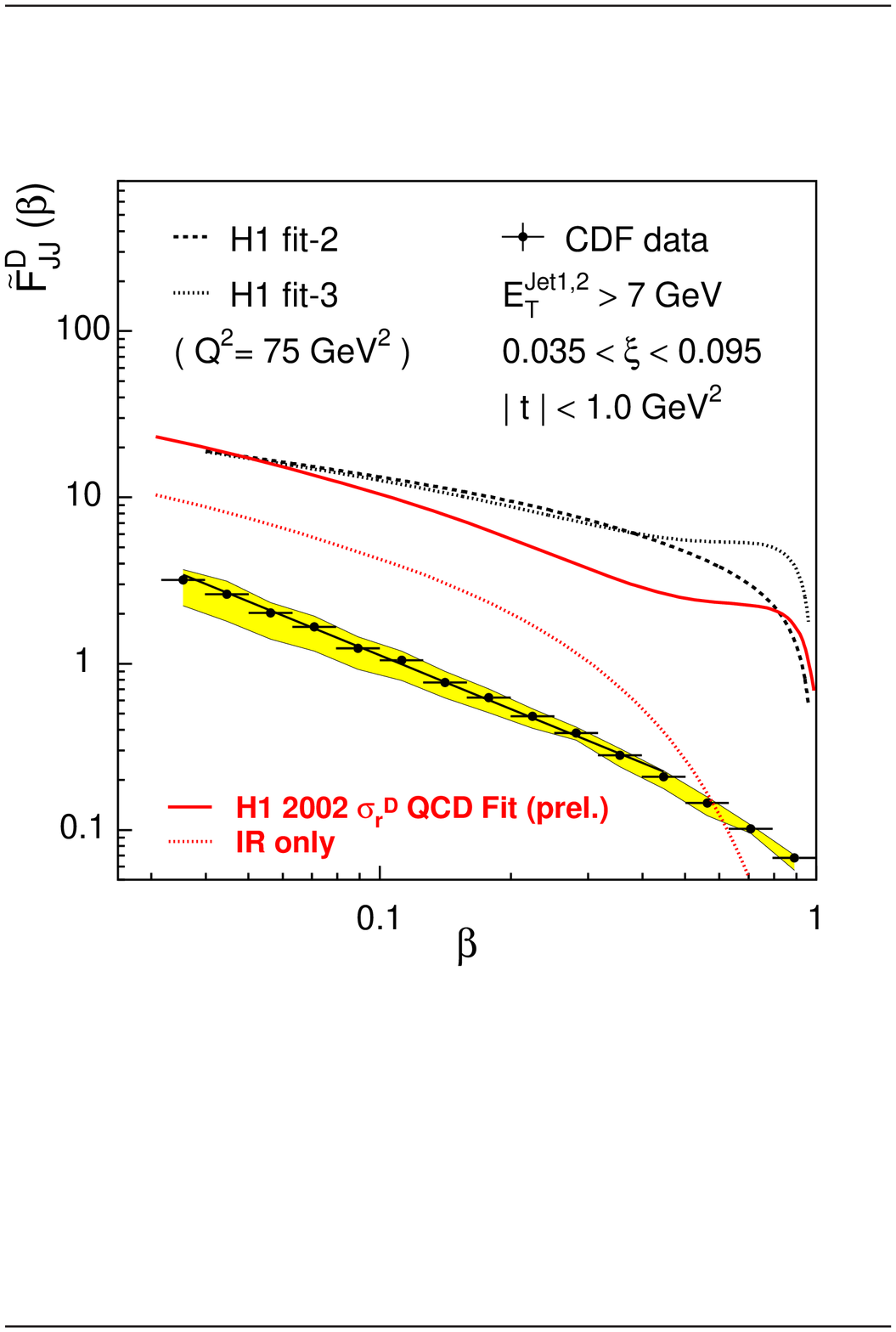,width=7cm,clip=true}
\vspace{6.cm}
\caption{Comparison between the CDF measurement of diffractive structure
function (black points) with the H1 diffractive PDFs.}
\label{cdfh}
\end{center}
\end{figure} 


Note that diffractive distributions are process-independent
functions.  They appear not only in inclusive diffraction but also in
other processes where diffractive hard-scattering factorization holds.  
The cross section of such a process can be
evaluated as the convolution of the relevant parton-level
cross section with the diffractive PDFs (DPDFs).
For instance, the cross section
for charm production in diffractive DIS can be calculated at leading order
in $\alpha_s$ from the $\gamma^* g \rightarrow c \bar c$ cross section and
the diffractive gluon distribution.  An analogous statement holds for jet
production in diffractive DIS. Both processes have been analyzed at
next-to-leading order in $\alpha_s$ and are found to be consistent with
the factorization theorem \cite{collins}.

A natural question to ask is whether one can use the DPDFs
extracted at HERA to describe hard diffractive processes such as the
production of jets, heavy quarks or weak gauge bosons in $p\bar{p}$
collisions at the Tevatron.  Fig.~\ref{cdfh} shows results on
diffractive dijet production from the CDF collaboration
compared to the expectations based on the 
DPDFs from HERA \cite{cdf}.  The discrepancy is spectacular:
the fraction of diffractive dijet events at CDF is a factor 3 to 10
smaller than would be expected on the basis of the HERA data. The same
type of discrepancy is consistently observed in all hard diffractive
processes in $p\bar{p}$ events.  In
general, while at HERA hard diffraction contributes a fraction of order
10\% to the total cross section, it contributes only about 1\% at the
Tevatron.
This observation of QCD-factorization breaking in hadron-hadron
scattering can be interpreted as a survival gap probability or a soft color interaction
which needs
to be considered in such reactions.
In fact, from a fundamental point of view, diffractive hard-scattering factorization does not apply to
hadron-hadron collisions.
Attempts to establish corresponding factorization theorems fail,
 because of interactions between spectator partons of the colliding
   hadrons.  The contribution of these interactions to the cross section
   does not decrease with the hard scale.  Since they are not associated
   with the hard-scattering subprocess, we no
   longer have factorization into a parton-level cross section and the
   parton densities of one of the colliding hadrons. These
interactions are generally soft, and we have at present to rely on
phenomenological models to quantify their effects \cite{cdf}. 
The yield of diffractive events in hadron-hadron collisions is then lowered
precisely because of these soft interactions between spectator partons
(often referred to as re-interactions or multiple scatterings).  
They can produce additional final-state particles which fill the would-be
rapidity gap (hence the often-used term rapidity gap survival).  When
such additional particles are produced, a very fast proton can no longer
appear in the final state because of energy conservation.  Diffractive
factorization breaking is thus intimately related to multiple scattering
in hadron-hadron collisions. Understanding and describing this
phenomenon is a challenge in the high-energy regime that will be reached
at the LHC \cite{afp}.
We can also remark simply that
the collision partners, in $pp$ or $p\bar{p}$ reactions, are both
composite systems of large transverse size, and it is not too
surprising that multiple interactions between their constituents can
be substantial.  In contrast, the virtual photon in $\gamma^* p$
collisions has small transverse size, which disfavors multiple
interactions and enables diffractive factorization to hold.  According
to our discussion, we may expect that for
decreasing virtuality $Q^2$ the photon behaves more and more like a
hadron, and diffractive factorization may again be broken.

\section{A brief comment on unitarity and diffraction}

\begin{figure}[htbp]
\centerline{\includegraphics[width=5.cm]{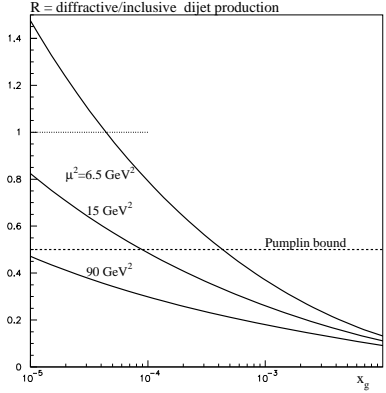}}
\caption{The ratio of diffractive to inclusive dijet production
cross section as a function of $x_{Bj}$ of the gluon for different scales
of the hard scattering, for the diffractive PDFs. Also shown is the
unitarity limit, called Pumplin bound.}
\label{fig:xgam2}
\end{figure}

Kaidalov et al.~\cite{kaidalov} have investigated what fraction of the gluon
distribution in the proton leads to diffractive final states. The
ratio of diffractive to inclusive dijet production cross sections as a
function of $x_{Bj}$ of the gluon, for different hard scattering scales and
for the diffractive PDFs is presented in Fig.~\ref{fig:xgam2}.
This ratio should be smaller than 0.5~\cite{pumplin}, while for scales
$\mu^2 =15$ GeV$^{2}$ this limit is exceeded for $x=10^{-4}$. This
indicates that unitarity effects may already be present in diffractive
scattering.

\section{The dipole picture of hadronic diffraction}
\begin{figure}[t]
   \vspace*{-1cm}
    \centerline{
     \epsfig{figure=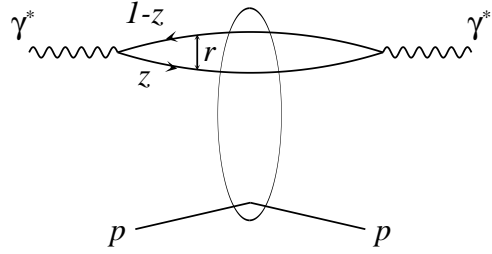,width=8cm}
               }
    \vspace*{-0.5cm}
\caption{Picture for the total
cross section ($\gamma^{*
}p\rightarrow\gamma^{*}p$) in the dipole model.
}
\label{f2dipoletot}
\end{figure}

The dynamics behind diffractive DIS can be easier understood if the
process is viewed in the rest frame of the proton. The virtual photon
develops a partonic fluctuations, whose lifetime is $\tau
=1/2m_px$~\cite{Ioffe:1984}. At the small $x_{Bj}$ typical of HERA, where
$\tau \sim 10 - 100$ fm, it is the partonic state rather than the
photon that scatters off the proton
(see Fig. \ref{f2dipoletot}). If the scattering is elastic, the
final state will have the features of diffraction.

The fluctuations of the $\gv$ are described by the wave functions of
the transversely and longitudinally polarized $\gv$ which are known
from perturbative QCD. Small and large partonic configurations of the
photon fluctuation are present. For large configurations
non-perturbative effects dominate in the interaction and the
treatment of this contribution is subject to modeling.  For a small
configuration of partons (large relative $k_T$) the total
interaction cross section of the created color dipole on a proton target can
be expressed as
\begin{eqnarray}
\sigma_{q\bar{q}p}&=&\frac{\pi^2}{3}r^2\alpha_S(\mu)xg(x,\mu) \, ,
\label{eq:qqp} \\
\sigma_{q\bar{q}gp}&\simeq& \sigma_{ggp}=\frac{9}{4}\sigma_{q\bar{q}p} \, ,
\label{eq:qqgp}
\end{eqnarray}
where $r$ is the transverse size of the color dipole and $\mu \sim
1/r^2$ is the scale at which the gluon distribution $g$ of the proton
is probed. The corresponding elastic cross section is obtained from
the optical theorem. In this picture, the gluon dominance in
diffraction results from the dynamics of perturbative QCD (see
equation~(\ref{eq:qqgp})) \cite{dipole}.

Models of diffraction that follow this approach are quite successful
in describing both the inclusive $F_2$ and the diffractive $F_2^D$
measurements, where the former are used to parameterize the
dipole-proton cross section \cite{dipole}. 

\section{Exclusive processes in DIS}

The presence of small size $q\bar{q}$ configurations in the photon can
be tested in exclusive vector meson (VM) production as well as for
deeply inelastic Compton scattering. At high energy (low $x_{Bj}$) and in the
presence of a large scale (large $Q^2$ or heavy flavor), these
reactions are expected to be driven by two-gluon exchange.

\begin{figure}[htbp]
\begin{center}
\includegraphics[width=0.8\hsize]{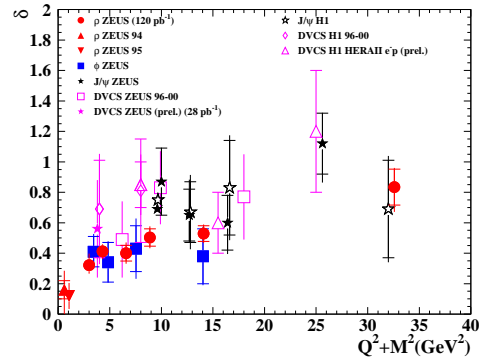}
\caption{ Logarithmic derivatives $\delta=d\log\sigma(\gvp)/d
\log W$ as a function of $Q^2+M_V^2$ for exclusive VM production. 
}
\label{fig:deltatop}
\end{center}
\end{figure}

A closer look at the theory of exclusive processes in QCD shows that
the two partons taking part in the exchange do not carry the same
fraction of the proton momentum. That makes these processes sensitive
to correlations between partons, which are encoded in the so-called
generalized parton distributions, GPDs \cite{gpds}. These new
functions relate in various limits to the parton distributions, form
factors and orbital angular momentum distributions. The motivation
behind studies of exclusive processes is to establish the region of
validity of pQCD expectations and ultimately to pursue a full mapping
of the proton structure, which cannot be achieved in inclusive
measurements \cite{Schoeffel:2009aa}.

\begin{figure}[htbp]
\begin{center}
\includegraphics[width=0.8\hsize]{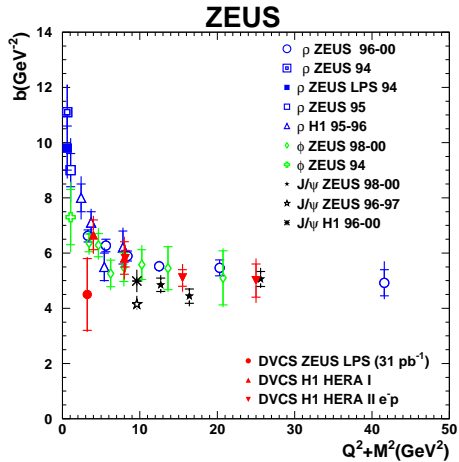}
\caption{ Exponential slope of the $t$ distribution measured for exclusive
VM production as a function of $Q^2+M_V^2$.
}
\label{fig:deltabottom}
\end{center}
\end{figure}

The cross section for the exclusive processes is expected to rise with
$W$, with the rate of growth increasing with the value of the hard
scale. A compilation of logarithmic derivatives
 $\delta=d\log\sigma(\gvp)/d\log W$, for $\rho$, $\phi$
and $J/\psi$ exclusive production, as a
function of the scale defined as $Q^2+M_{V}^2$, where $M_{V}$ is
the mass of the VM, is presented in Fig.~\ref{fig:deltatop} \cite{Schoeffel:2009aa}.

In Fig.~\ref{fig:deltatop}, we observe a
universal behavior, showing an increase of $\delta$ as the scale
becomes larger. The value of $\delta$ at low scale is the one expected
from the soft Pomeron intercept, while the one at large
scale is in accordance with twice the logarithmic derivative of the
gluon density with respect to $W$.
Then, when $\delta$ is measured to be of the order of $0.6$ or higher, the
process is hard and calculable in perturbative QCD.

Another fundamental measurement concerns the cross section 
of exclusive VM production, differential in $t$, where
 $t=(p-p')^2$ is the momentum transfer (squared) at the proton vertex.
A parametrization in $d\sigma/dt \sim e^{-b|t|}$
gives a  very good description of all measurements in the kinematic
range of HERA (at low $x<0.01$).
Then, when $Q^2+M_V^2$ is increasing,
which corresponds to a decreasing transverse size of the $q\bar{q}$ dipole, 
the $t$ distribution is
expected to become universal, independent of the scale and of the
VM. The exponential slope of the $t$ distribution, $b$, reflects then
the size of the proton. A
compilation of
measured $b$ values is presented in Fig.\ref{fig:deltabottom}.
Around $Q^2+M_V^2$ of about $15$ GeV$^2$ indeed the $b$ values become
universal.
A qualitative understanding of this behavior is simple.
Indeed, $b$ is essentially the sum of a component
coming from the probe in $1/\sqrt{Q^2+M_{VM}^2}$ and a 
component related to the target nucleon.
Then, at large $Q^2$ or large $M_{VM}^2$, the $b$ values
decrease to the solely target component.
That's why in Fig.~\ref{fig:deltabottom}, we observe that for
large $Q^2$ or for heavy VMs, like $J/\psi$, $b$ is reaching a 
universal value of about $5$ GeV$^{-2}$, scaling with $Q^2$
asymptotically. 
This value is related to the size of the target probed
during the interaction  and we do not 
expect further decrease of $b$ when increasing the scale,
once a certain scale is reached \cite{Schoeffel:2009aa}.

\section{Nucleon Tomography}

Measurements of the $t$-slope parameters $b$, presented in the previous section
(Fig. ~\ref{fig:deltabottom}),
are key measurements for almost all exclusive processes.
 Indeed,
a Fourier transform from momentum
to impact parameter space readily shows that the $t$-slope $b$ is related to the
typical transverse distance between the colliding objects.
At high scale, the $q\bar{q}$ dipole is almost
point-like, and the $t$ dependence of the cross section is given by the transverse extension 
of the gluons (or sea quarks) in the  proton for a given $x_{Bj}$ range.
In particular for DVCS \cite{dvcs}, interpretation of $t$-slope measurements
does not suffer from the lack of knowledge of the VM wave function.
Then, a DVCS cross section, differential in $t$, is directly related to 
GPDs \cite{gpds}
More precisely, from GPDs, we can compute
a parton density which also depends on a spatial degree of freedom, the transverse size (or impact parameter), labeled $R_\perp$,
in the proton. Both functions are related by a Fourier transform 
$$
PDF (x, R_\perp; Q^2) 
\;\; \equiv 
$$
$$
\int \frac{d^2 \Delta_\perp}{(2 \pi)^2}
\; e^{i ({\Delta}_\perp {R_\perp})} 
\; GPD (x, t = -{\Delta}_\perp^2; Q^2).
$$
Thus, the transverse extension $\langle r_T^2 \rangle$
 of gluons (or sea quarks) in the proton can be written as
$$
\langle r_T^2 \rangle
\;\; \equiv \;\; \frac{\int d^2 R_\perp \; PDF(x, R_\perp) \; R_\perp^2}
{\int d^2 R_\perp \; PDF(x, R_\perp)} 
$$
$$
= \;\; 4 \; \frac{\partial}{\partial t}
\left[ \frac{GPD (x, t)}{GPD (x, 0)} \right]_{t = 0} = 2 b
$$
where $b$ is the exponential $t$-slope.
Measurements of  $b$
presented in Fig. \ref{fig:deltabottom}
corresponds to $\sqrt{r_T^2} = 0.65 \pm 0.02$~fm at large scale $Q^2$ for $x_{Bj} < 10^{-2}$.
This value is smaller that the size of a single proton, and, in contrast to hadron-hadron scattering, it does not expand as energy $W$ increases.
This result is consistent with perturbative QCD calculations in terms of a radiation cloud of gluons and quarks
emitted around the incoming virtual photon. In short, gluons are located at the preiphery of the proton as measured 
here and valence quarks are assumed to form the core of the proton at small value of $\sqrt{r_T^2}$.

In other words,  the
  Fourier transform of the DVCS amplitude is the amplitude to find quarks at
$R_\perp$ in an image plane after focusing by an idealized lens.
The square of the profile amplitude, producing the PDF (in transverse plane)
is positive, real-valued, and corresponds to the
image, a weighted probability to find quarks in the transverse
image plane.  

\section{Perspectives at CERN}
The complete parton imaging in the nucleon would need to get  measurements of $b$ for
several values of $x_{Bj}$, from the low $x_{Bj} < 0.01$ till $x_{Bj}>0.1$. Experimentally,
it appears to be impossible. Is it the breakout of quark and gluon imaging in the proton?
In fact, there is one way to recover $x_{Bj}$ and $t$ correlations over the whole $x_{Bj}$
domain: we need to measure a Beam Charge Asymmetry (BCA).

A determination of a cross section asymmetry with respect to the beam
charge has been realized by the H1 experiment by measuring the ratio
$(d\sigma^+ -d\sigma^-)/ (d\sigma^+ + d\sigma^-)$ as a function of $\phi$,
where $\phi$ is the azimuthal angle between leptons and proton plane.
The result has recently been obtained by the H1 collaboration 
(see Ref. \cite{dvcs}) with  a fit in $\cos \phi$.
After applying a deconvolution method to account for the  resolution on $\phi$,
the coefficient of the $\cos \phi$ dependence is found to be $p_1 = 0.16 \pm 0.04 (stat.) \pm 0.06 (sys.)$.
This result represents obviously a major progress in the understanding of the very recent field of the 
parton imaging in the proton. We are at the hedge of the giving a new reading on the most fundamental question to know
how the proton is built up by quarks and gluons.

Feasibilities for future BCA measurements at COMPASS have been studied extensively
in the last decade \cite{dhose}. COMPASS is a fixed target experiment which can use
100 GeV muon beams and hydrogen targets, and then access experimentally the DVCS process $\mu p \rightarrow \mu \gamma p$.
The BCA can be determined when using positive and negative muon beams.
One major interest is the kinematic coverage from $2$ GeV$^2$ till $6$ GeV$^2$ in $Q^2$
and  $x_{Bj}$ ranging from $0.05$ till $0.1$. It means that it is possible to avoid
the kinematic domain dominated by higher-twists and non-perturbative effects 
(for $Q^2 < 1$ GeV$^2$) and keeping a
$x_{Bj}$ range which is extending the HERA (H1/ZEUS) domain.

\section{Conclusions}

We have reviewed the most recent experimental results from
hard diffractive scattering at HERA and Tevatron.
We have shown that
many aspects of diffraction in $ep$ collisions can be successfully 
described in QCD if a hard scale is
present. A key to this success are factorization theorems, which render parts of the dynamics 
accessible to calculation in perturbation theory. The remaining 
non-perturbative quantities, namely diffractive
PDFs and generalized parton distributions, can be extracted from 
measurements and contain specific
information about small-$x_{Bj}$ partons in the 
proton that can only be obtained in diffractive processes. To
describe hard diffractive hadron-hadron collisions is more 
challenging since factorization is broken by
re-scattering between spectator partons. 
These re-scattering effects are of interest in their own right 
because of their intimate relation with multiple scattering effects, which at 
LHC energies are expected to be
crucial for understanding the structure of events in hard collisions. 

A combination of data on inclusive
and diffractive $ep$ scattering hints at the onset of parton saturation at 
HERA, and the phenomenology
developed there is a helpful step towards understanding high-density 
effects in hadron-hadron collisions.
In this respect,
we have discussed  
a very important aspect that makes diffraction in DIS  so
interesting at low $x_{Bj}$. Its interpretation in the
dipole formalism and its connection
to saturation effects. Indeed, diffraction in DIS has appeared as
a well suited process to analyze saturation effects 
at large gluon density in the proton. In the dipole model,
it takes a simple and luminous form, with the introduction of
the so-called saturation scale $Q_s$. Diffraction is then dominated
by dipoles of size $r \sim 1/Q_s$. In particular, it provides a simple 
 explanation of the
constance of the ratio of diffractive to total cross sections
as a function of $W$ (at fixed $Q^2$ values).

Then, exclusive processes in DIS, like VMs production or DVCS, have appeared
as key reactions to trigger the generic mechanism of diffractive scattering.
Decisive measurements have been performed recently, in particular 
concerning  dependences of exclusive processes
cross section within the momentum exchange (squared) at the proton vertex, $t$.
This allows to extract first  experimental features concerning proton 
tomography, on how partons are localized in the proton.
It provides 
a completely new information on the spatial extension of partons
inside the proton (or more generally hadrons), as well as on the 
correlations of longitudinal momenta.
A unified picture of this physics is encoded in the GPDs formalism.
We have shown that
Jefferson laboratory experiments or prospects at COMPASS are essential,
to gain relevant information on GPDs.
Of course, we do not forget that the dependence of GPDs on three
kinematical variables, and the number of distributions describing
different helicity combinations present a considerable complexity.  In
a sense this is the price to pay for the amount of physics information
encoded in these quantities.  It is however crucial to realize that
for many important aspects we need not fully disentangle this
complexity.  The relation of longitudinal and transverse structure
of partons in a nucleon, or of nucleons in a nucleus, can be studied
quantitatively from the distribution in the two external kinematical
variables $x_{Bj}$ and $t$.



\begin{thebibliography}{10}

\bibitem{f2d97}
A.~Aktas {\it et al.}  [H1 Collaboration],
  Eur.\ Phys.\ J.\  C {\bf 48} (2006) 715;
  Eur.\ Phys.\ J.\  C {\bf 48} (2006) 749;
  ZEUS Collab., S.~Chekanov { et al.},
  Nucl.\ Phys.\ B {\bf 713} (2005) 3;
  Eur.\ Phys.\ J.\  C {\bf  38} (2004) 43.

\bibitem{marta}
M.~Ruspa et al.  [ZEUS Collaboration],
  [arXiv:hep-ex/0808.0833].

\bibitem{collins} 
J.C.~Collins,
  Phys.\ Rev.\  D {\bf 57} (1998) 3051
  [Erratum-ibid.\  D {\bf 61} (2000) 019902].

\bibitem{owens} 
  J.F. Owens, {\ Phys. Rev. } D {\bf 30} (1984) 943.

\bibitem{lolo1} 
C.~Royon, L.~Schoeffel, S.~Sapeta, R.B.~Peschanski and E.~Sauvan,
  Nucl.\ Phys.\  B {\bf 781} (2007) 1;
C.~Royon, L.~Schoeffel, R.B.~Peschanski and E.~Sauvan,
  Nucl.\ Phys.\  B {\bf 746} (2006) 15;
C.~Royon, L.~Schoeffel, J.~Bartels, H.~Jung and R.~B.~Peschanski,
  Phys.\ Rev.\  D {\bf 63} (2001) 074004.

\bibitem{zeus:diffqcd} ZEUS Collaboration, 
{\it `A QCD analysis of diffractive DIS 
data from ZEUS'} [ZEUS-pub-09-010]. 

\bibitem{cdf} 
T.~Affolder {\it et al.}  [CDF Collaboration],
Phys.\ Rev.\ Lett.\  {\bf 84} (2000) 5043.

\bibitem{plb}
  C.~Marquet and L.~Schoeffel,
  Phys.\ Lett.\  B {\bf 639} (2006) 471.

\bibitem{afp} AFP TDR in ATLAS to be submitted; see: 
http://project-rp220. web.cern.ch/project-rp220/index.html. 


\bibitem{kaidalov}
A.~B.~Kaidalov, V.~A.~Khoze, A.~D.~Martin and M.~G.~Ryskin,
Phys.\ Lett.\ B {\bf 567}, 61 (2003), hep-ph/0306134.

\bibitem{pumplin}
J.~Pumplin,
Phys.\ Rev.\ D {\bf 8} (1973) 2899.


\bibitem{Ioffe:1984}
B.~L.~Ioffe, V.~A.~Khoze and L.~N.~Lipatov,
``Hard Processes. Vol. 1: Phenomenology, Quark Parton Model,''
{\it  Amsterdam, Netherlands: North-Holland, 1984}.
\bibitem{dipole}
A.H. Mueller, { Nucl. Phys.} {\bf   B335} (1990) 115;  
%
N.N. Nikolaev and B.G. Zakharov, {Zeit. f\"ur. Phys.} {\bf   C49} (1991) 607;
A. Bialas and R. Peschanski, {Phys. Lett.} {\bf   B378} (1996) 302; 
 { Phys. Lett.} {\bf   B387} (1996) 405; 
%
H.~Navelet, R.~Peschanski, C.~Royon, S.~Wallon,
Phys. Lett. B {\bf 385} (1996) 357;  
%
H.~Navelet, R.~Peschanski, C.~Royon, Phys. Lett. B366 (1996) 329; 
%
K.~J.~Golec-Biernat and M.~Wusthoff,
Phys.\ Rev.\  D {\bf 59} (1999) 014017;  
Phys.\ Rev.\  D {\bf 60} (1999) 114023; 
A.~Bialas, R.~Peschanski, C. Royon,
Phys. Rev. D {\bf 57} (1998) 6899;  
%
S.~Munier, R.~Peschanski, C.~Royon, 
Nucl. Phys. B {\bf 534} (1998) 297; 
%
 E.~Iancu, K.~Itakura and S.~Munier,
  Phys.\ Lett.\ B {\bf 590} (2004) 199.

\bibitem{gpds}
M.~Diehl, T.~Gousset, B.~Pire and J.~P.~Ralston,
Phys.\ Lett.\  B {\bf 411} (1997) 193;  
%
L.~L.~Frankfurt, A.~Freund and M.~Strikman,
Phys.\ Rev.\  D {\bf 58} (1998) 114001
[Erratum-ibid.\  D {\bf 59} (1999) 119901];   
%
A.~V.~Belitsky, D.~Mueller and A.~Kirchner,
Nucl.\ Phys.\  B {\bf 629} (2002) 323; 
%
M.~Diehl,
{Phys.\ Rept.}\ {\bf 388}, 41 (2003).

\bibitem{Schoeffel:2009aa}
  L.~Schoeffel,
  arXiv:0908.3287 [hep-ph].

\bibitem{dvcs}
  H.~Collaboration,
  arXiv:0907.5289 [hep-ex]; Eur.\ Phys.\ J.\  C {\bf 44} (2005) 1;
  S.~Chekanov {\it et al.}  [ZEUS Collaboration],
  JHEP {\bf 0905} (2009) 108.



\bibitem{dhose}
 N.~d'Hose et al.,
Nucl.\ Phys.\  A {\bf 711} (2002) 160.

\end{thebibliography}
\end{document}